\documentclass{article}

\usepackage{mathtools}
\usepackage{graphicx} 
\usepackage{amsmath}
\usepackage{amsfonts}
\usepackage{amssymb}
\usepackage[margin=1in]{geometry}
\usepackage{natbib}
\usepackage{amsthm}
\usepackage{authblk}
\usepackage{hyperref}
\usepackage{setspace}
\hypersetup{
colorlinks=true,
linkcolor=red,
filecolor=magenta,
citecolor=blue,
urlcolor=blue,
}
\usepackage{appendix}

\newtheorem{theorem}{Theorem}
\newtheorem{assumption}{Assumption}

\mathtoolsset{showonlyrefs}

\newcommand{\Pw}{\lambda}
\newcommand{\g}{\pi}
\newcommand{\Qw}{\tilde{\mu}}
\newcommand{\Qa}{\dot{\mu}}
\newcommand{\Qaw}{\bar{\mu}}
\newcommand{\indep}{\mbox{$\perp\!\!\!\perp$}} 
\newcommand{\excessrisk}{\mathrm{D}} 
\newcommand{\smr}{\mathrm{R}} 

\newcommand{\tmle}{\mathrm{tmle}}
\newcommand{\logit}{\mathrm{logit}}

\newcommand{\E}{\mathsf{E}}
\newcommand{\direct}{\phi}
\newcommand{\indirect}{\psi}
\newcommand{\R}{\mathsf{R}}
\newcommand{\D}{\mathsf{D}}

\newcommand{\arxiv}{1}

\ifx\arxiv\undefined 
    \date{}
\else
    \date{}
\fi

\title{Doubly Robust Nonparametric Efficient Estimation for Provider Evaluation}
\author[1,*]{Herbert Susmann}
\author[2]{Yiting Li}
\author[2]{Mara A. McAdams-DeMarco}
\author[1]{Iv\'an D\'iaz}
\author[1]{Wenbo Wu}

\affil[1]{\small Division of Biostatistics, Department of Population
  Health, NYU Grossman School of Medicine, USA}
\affil[2]{\small Department of Surgery, NYU Grossman School of Medicine, USA}
\affil[*]{\small Corresponding author: susmah01@nyu.edu}

\begin{document}

\maketitle

\abstract{
Provider profiling has the goal of identifying healthcare providers with exceptional patient outcomes. When evaluating providers, adjustment is necessary to control for differences in case-mix between different providers. Direct and indirect standardization are two popular risk adjustment methods. In causal terms, direct standardization examines a counterfactual in which the entire target population is treated by one provider. Indirect standardization, commonly expressed as a standardized outcome ratio, examines the counterfactual in which the population treated by a provider had instead been randomly assigned to another provider. Our first contribution is to present nonparametric efficiency bound for direct and indirectly standardized provider metrics by deriving their efficient influence functions. Our second contribution is to propose fully nonparametric estimators based on targeted minimum loss-based estimation that achieve the efficiency bounds. The finite-sample performance of the estimator is investigated through simulation studies. We apply our methods to evaluate dialysis facilities in New York State in terms of unplanned readmission rates using a large Medicare claims dataset. A software implementation of our methods is available in the \texttt{R} package \texttt{TargetedRisk}.}
\\[1em]
\begin{keywords}
direct standardization; indirect standardization; nonparametric statistics; provider profiling; targeted minimum loss-based estimation
\end{keywords}

\section{Introduction}
Provider profiling is concerned with quantifying the performance of healthcare providers with respect to patient outcomes \citep{normand1997profiling}. In spite of the extensive literature on regression-based models, provider profiling can naturally be framed as a causal inference problem in which the central concern is how patient outcomes would have been different if patients had been treated by a provider other than the one by which they were actually treated \citep{longford2019performance}. As the randomization of patients to providers is not possible in a profiling context where patient outcomes are observed without any intervention, causal inference methods for observational data can be applied to address biases resulting from non-randomization.

A core methodological issue of profiling analysis lies in adjusting for differences in case-mix between providers \citep{zaslavsky2001, shahian2008}.
Direct and indirect standardization are two of the principal methods for case-mix adjustment \citep{keiding2014standardization}. In a causal interpretation, direct standardization is based on counterfactual outcomes had all patients in the population been treated by a given provider. Directly standardized outcomes can be directly compared across providers as they are based on a common index population. In practice, this method can be challenging when there are significant differences in case-mix across providers. If a provider rarely or never treats a particular subset of patients from the index population, it is difficult or even impossible to accurately estimate the counterfactual outcomes for this subpopulation by the provider. Estimates of provider effects in this region could be based on model extrapolation, which may be significantly biased \citep{varewyck2014shrinkage, silva2022}. In casual terms, this is referred to as a practical violation of the positivity assumption: there exists a stratum of covariates with little or no probability of being treated by a particular provider.

Indirect standardization, on the other hand, compares the outcomes of patients treated by given provider to their counterfactual outcomes had they been randomly reassigned to another provider according to a specified probability distribution. There is some flexibility in defining the counterfactual reassignment mechanism: patients could be assigned to a provider uniformly at random, to a provider with ``average'' performance \citep{he2013}, to a hypothetical provider formed by pooling patients across all providers or consolidating providers within a certain geographical region \citep{varewyck2014shrinkage, han2023privacypreserving}, or to a provider that tends to treat patients with similar covariates. By carefully choosing the randomization process, we can avoid the positivity problem inherent in direct standardization \citep{daignault2017doublerobust}. However, drawback is that indirectly standardized outcomes for a provider depend on the provider's case-mix, making it difficult to compare across providers \citep{shahian2020misuse}.

In this article, we formalize the standardization problem in nonparametric causal terms using structural equation models \citep{pearl2009}, drawing on pathbreaking work by \cite{daignault2017doublerobust}. Direct standardization under this framework amounts to estimating a counterfactual outcome with a categorical exposure. The corresponding target parameter is thus identifiable under standard causal assumptions. For indirect standardization, we consider a randomization process in which patients have equal probabilities of being assigned to the counterfactual ``average'' provider, effectively weighting all actual providers by their patient volumes. The indirectly standardized outcome ratio is then interpretable as the ratio of a provider's mean patient outcome to that of a reference provider at the overall average level of care. Causal identification results unveil that under this randomization process the indirectly standardized outcome ratio can be estimated without modeling any provider-covariate interactions \citep{daignault2017doublerobust}.

Numerous statistical approaches have been developed for risk adjustment in provider profiling, with regression-based hierarchical random- and fixed-effects models being the mostly commonly used \citep[e.g.,][]{normand1997profiling, he2013, lee2016hierarchical, estes2018time, wu2022improving, haneuse2022measuring}; there is a large literature comparing the relative benefits of fixed vs random provider effects \citep{alma2013riskadjustment, kalbfleisch2013monitoring, kalbfleisch2018discussion, hansen2023review}. In addition, causal inference approaches have been applied such as matching \citep{silber2014matching, longford2019performance}, inverse probability of treatment weighting \citep{tang2020ipw}, prognostic score weighting \citep{lee2022facility, lee2024prognostic}, and balancing weights \citep{keele2023balancing}. 

Doubly robust estimators of standardization parameters have also been considered for profiling providers in various contexts \citep{varewyck2014shrinkage, spertus2016direct, daignault2017doublerobust, han2023privacypreserving}. These estimators are consistent if either of the two nuisance parameters -- the provider assignment mechanism or the outcome model -- is estimated consistently. However, extant doubly robust estimators rely on parametric models for both nuisance parameters. For instance, \citet{spertus2016direct} utilized TMLE to estimate directly standardized excess mortality, employing a logistic regression as the outcome model and multinomial propensity scores to characterize the provider assignment mechanism. In most real-world profiling applications, such parametric modeling assumptions are unlikely to hold, leading to inconsistency in the estimators. 

Our work fills two imporant gaps in the extant research. First, we present nonparametric efficiency bounds for both direct and indirect standardization parameters; that is, we find the best asymptotic variance possible for regular estimators of the parameters. These efficiency bounds are defined as the variance of the parameters \textit{efficient influence functions} (EIFs). The EIF of the direct standardization parameter has previously been derived. Our primary contribution is to derive the EIF for indirect standardization parameters, which is novel. Knowledge of the efficiency bounds via the EIFs yields key insights into the relative difficulty in estimating direct vs indirect parameters. In addition, the development of this nonparametric efficiency theory for the standardization parameters, including their efficient influence functions, paves the way for valid statistical inference including formation of confidence intervals and hypothesis testing \citep{Bickel97}.

The second research gap we address is that existing doubly robust standardization estimators are based on estimating nuisance parameters using parametric models, which in real-world scenarios are likely inconsistent. We develop nonparametric doubly robust estimators based on targeted minimum loss-based estimation (TMLE) that do not require any such restrictive parametric assumptions \citep{vanderLaanRose11}. Nuisance parameters are estimated using flexible machine learning algorithms or ensembles thereof. Remarkably, the TMLE estimators provably attain the nonparametric efficiency bounds defined by the EIFs of the standardization parameters even when nuisance parameters are estimated using data-adaptive algorithms. Our main novel contribution is to develop a novel estimator based on TMLE for the indirect standardization parameter. We also present a TMLE approach for direct standardization, which was previously considered in a more general causal inference context \citep{vanderlaan2006tmle}. Our inclusion of TMLE estimators for both indirect and direct standardization parameters makes this article a complete account of nonparametric doubly robust TMLE for risk-standardized provider profiling.

The rest of the paper unfolds as follows. In Section~\ref{section:causal} we introduce the causal model and give causal identification results for direct and indirect standardization parameters. Section~\ref{section:efficiency-theory} investigates the semiparametric efficiency properties of the parameters, and Section~\ref{section:estimation} builds on these results to form targeted nonparametric efficient estimators. Section~\ref{section:simulations} presents results from simulation studies. We illustrate our methods in Section~\ref{section:application} by estimating direct and indirect standardization measures for dialysis providers for patients with end-stage renal disease, where the outcome of interest is unplanned hospital readmission within 30 days of an index discharge.

\section{Causal Framework}
\label{section:causal}
Suppose we observe a sample $o_1, \dots, o_n$ where each observation is a draw of the generic variable $O = (W, A, Y)$ with $W$ a vector of patient covariates, $A \in \{ 1, \dots, m \}$ indicating the provider, and $Y$ a binary ($Y \in \{0, 1 \}$) or continuous ($Y \subseteq \mathbb{R}$) outcome. We assume that $O$ is drawn independently from a law $P_0$. We assume only that $P_0$ falls in the nonparametric model $\mathcal{M}$, the set of all possible laws defined on the support $\mathcal{O}$ of $O$. 

\paragraph{Notation} We will commonly make reference to specific parts of a distribution $P \in \mathcal{M}$. For convenience, let $\Pw_P(W)$ be the marginal distribution of $W$ under $P$, and define
\begin{align}
  \g_P(a, w)       &= P(A = a \mid W = w), \\
  \Qaw_P(a, w) &= \E_P[Y \mid A = a, W = w], \\
  \Qa_P(a)    &=  \E_P[Y \mid A = a], \\
  \Qw_P(w)    &=  \E_P[Y \mid W = w].
\end{align}
Note that $\mu$ always refers to a conditional expectation of $Y$ with respect to some set of variables. The parameter $\g_P$ is referred to as the ``propensity score" or ``probability of treatment" model and $\Qaw_P$ is referred to as the ``outcome model". 
For convenience, for a parameter $\theta_P$ indexed by $P \in \mathcal{M}$ we will write $\theta_0 := \theta_{P_0}$ to refer to its value under the true data generating distribution $P_0$. For a distribution $P \in \mathcal{M}$, we will write $\E_P$ to denote an expectation with respect to $P$, and $\E_0$ to denote expectation with respect to $P_0$. We may also write $Pf = \int f dP$ and $P_n f = n^{-1} \sum_{i=1}^n f(o_i)$. 

\subsection{Causal Model}
We define the causal structure of the problem within the structural causal model framework \citep{pearl2009}, noting that it would also be possible to equivalently define the problem using a different causal formalism such as the potential outcomes framework or single-world intervention graphs \citep{rubin1974, richardson2013single}.
We assume the data are generated according to the following nonparametric structural model:
\begin{align}
    W &= f_W(U_W), \\
    A &= f_A(W, U_A), \\
    Y &= f_Y(W, A, U_Y),
\end{align}
where $f_W$, $f_A$, and $f_Y$ are arbitrary deterministic functions and $U_W$, $U_A$, and $U_Y$ are exogenous variables.
Let $Y^Z$ be the counterfactual outcome induced by intervening on the structural model and setting $A = Z$. That is, $Y^Z$ is the counterfactual outcome had the individual been (possibly contrary to fact) treated by provider $Z$. In other words, $Y^Z = f_Y(W, Z, U_Y)$. The observed outcome $Y$ corresponds to the counterfactual outcome $Y^Z$ when $Z = A$, the provider that in reality treated the patient. The counterfactual outcomes for all other providers are unobserved. We will consider stochastic and deterministic $Z$ assignments.

\subsection{Causal Estimands}
Next, we define a set of target causal estimands. Each estimand is an average of the counterfactual outcomes $Y^Z$ under different ways of counterfactually reassigning patients to providers. 
We focus on two causal estimands referred to as \textit{direct} and \textit{indirect} standardization parameters. The direct standardization parameter examines a counterfactual in which all patients in the population are reassigned to one provider. The indirect parameter adopts a stochastic intervention in which patients are randomly reassigned to a provider. To differentiate between the two in terms of notation, we will use $\direct$ for directly standardized parameters and $\indirect$ for indirectly standardized parameters.

\paragraph{Direct standardization}
For a provider $a' \in \{1, \dots, m \}$, the directly standardized outcome is defined as the expected outcome if every patient in the population had, possibly contrary to fact, been treated by provider $a'$:
\begin{align}
    \direct(P)(a') &= \E_P[Y^{a'}]. 
\end{align}
The direct standardization parameter can be interpreted as an average treatment effect for a categorical treatment variable.

We focus primarily on estimating the parameters $\direct(P)(a')$ for each $a'$. As each of these parameters is based on the same reference population, the directly standardized risk of each provider can be directly compared to one another, for example in the form of a league table. Alternatively, the parameter $\direct(P)(a')$ could be compared to a benchmark such as the median of the provider direct standardization parameters $\mathrm{median}\{ \direct(P)(a') : a' \in \{ 1, \dots, m \} \}$.

\paragraph{Indirect standardization}
Indirect standardization can be defined in terms of a stochastic intervention. Define a stochastic intervention by setting $A = Z$, where $Z$ is a draw from some conditional distribution $P_{Z | W}$. We focus on the setting where $P_{Z|W} = P_{A|W}$, following  \cite{daignault2017doublerobust}. This implies an intervention in which individuals are randomly assigned to another provider that tends to treat individuals with the same characteristics. 
For a provider $a' \in \{1, \dots, m \}$, define the expected outcome conditional on having been treated by that provider:
\begin{align}
    \psi_1(P)(a') &= \E_P[Y^{a'} | A = a']
\end{align}
Note that $\psi_1(P)(a')$ is not a counterfactual quantity, as it is simply the average of the observed outcomes of patients who were treated by provider $a'$. 
Next, define the expected counterfactual outcome under the stochastic intervention among individuals who were treated at provider $a'$:
\begin{align}
    \psi_2(P)(a') &= \E_P[Y^Z | A = a']. 
\end{align}
This is a truly counterfactual parameter, as it involves unobserved counterfactuals: what would have happened if patients who saw provider $a'$ had been randomly assigned a provider according to the law $P_{A | W}$? 

The quality of a provider can subsequently be evaluated by comparing $\psi_1(P)(a')$ and $\psi_2(P)(a')$. Intuitively, we want to evaluate whether patient outcomes for those who saw provider $a'$ would have been better if they had been randomly assigned to a provider who treats similar patients to provider $a'$ (this is $\psi_2(P)(a')$) compared to the outcomes after being treated by $a'$ (this is $\psi_1(P)(a')$). We consider two possible ways of comparing $\psi_1(P)(a')$ and $\psi_2(P)(a')$, via their ratio and difference. The indirectly standardized excess risk is defined as the difference of $\psi_1$ and $\psi_2$:
\begin{align}
    \psi_{\excessrisk}(P)(a') &= \psi_1(P)(a') - \psi_2(P)(a'),
\end{align}
where the subscript $\excessrisk$ is to emphasize that this parameter represents a \textit{difference}.
The indirectly standardized outcome ratio is defined as the ratio of $\psi_1$ to $\psi_2$: 
\begin{align}
    \psi_{\smr}(P)(a') &= \frac{\psi_1(P)(a')}{\psi_2(P)(a')},
\end{align}
where the subscript $\smr$ is in reference to the parameter's definition as a \textit{ratio}.

\subsection{Causal Identification}
The target causal estimands defined in the previous section depend on counterfactual variables which are not observed in practice. Causal identification refers to establishing the assumptions necessary to show that the causal parameters defined in terms of the counterfactuals $Y^Z$ can be expressed only in terms of the observed data. In order to identify the direct standardization parameters, the following assumptions are necessary:
\begin{assumption}[Exchangeability]
    \label{assumption:exchangeability}
    The counterfactual outcomes are conditionally independent of $A$ given $W$: for each $a' \in \{1, \dots, m\}$, it holds that $Y^{a'} \indep A \mid W$. 
\end{assumption}
\begin{assumption}[Positivity]
    \label{assumption:positivity}
    It holds $P_0$-almost surely that for any $a' \in \{1, \dots, m \}$, $\g(a', W) > 0$. 
\end{assumption}
Assumption~\ref{assumption:exchangeability} is a standard assumption that requires all confounders to be measured and included in $W$. Assumption~\ref{assumption:positivity} is also standard and requires that all strata of patient covariates have positive probability of being assigned to any hospital. However, in practical provider profiling scenarios Assumption~\ref{assumption:positivity} may be violated if there are providers who do not treat some strata of patients. 
Note that the consistency assumption required when using the potential outcomes framework (as in \citealt{daignault2017doublerobust}) is automatically satisfied by the structural causal model and is therefore not necessary. For the directly standardized outcome, Assumptions~\ref{assumption:exchangeability} and \ref{assumption:positivity} are sufficient to identify the parameter via the well-known g-computation identification result \citep{robins1986}:
\begin{align}
    \phi(P)(a') &= \E_P[\E_P[Y \mid A = a', W = w]]. 
\end{align}

To identify the indirect standardization parameters, we require Assumption~\ref{assumption:exchangeability} as before, but we require a weaker form of positivity:
\begin{assumption}[Weak positivity]
    \label{assumption:weak-positivity}
    It holds that for any $a' \in \{1, \ldots, m \}$, $P(A = a') > 0$. 
\end{assumption}
Assumptions~\ref{assumption:exchangeability} and \ref{assumption:weak-positivity} are sufficient to prove the identification result
\begin{align}
    \psi_1(P)(a') &= \E_P[Y \mid A = a'], \\
    &= \Qa_P(a'), \\
    \psi_2(P)(a') &= \E_P\left[ \E_P[Y \mid W] \mid A = a' \right] \\
    &= \E_P[\Qw_P(W) \mid A = a'].
\end{align}
Assumption~\ref{assumption:weak-positivity} is required so that the conditional expectations in $\indirect_1(P)(a')$ and $\indirect_2(P)(a')$ are well-defined for all $a'$. 
Identification of the indirect standardization parameter was originally proved in \cite{daignault2017doublerobust} under the stronger positivity Assumption~\ref{assumption:positivity}; however, it is not necessary, so we instead adopt the substantially weaker Assumption~\ref{assumption:weak-positivity}. For completeness, we include a proof of the identification result for $\indirect_2(P)(a')$ in the Supplemental Material using our notation, following the proof of \cite{daignault2017doublerobust}. 

As they are simple functions of $\indirect_1(P)(a')$ and $\indirect_2(P)(a')$, identification of $\indirect_{\excessrisk}(P)(a')$ and $\indirect_{\smr}(P)(a')$ follows straightforwardly under the same assumptions. Note that $\psi_2(P)(a')$ depends only on $\E_P[Y \mid W]$, rather than $\E_P[Y \mid A , W]$, as is required for the directly standardized outcome. In other words, it is not necessary to estimate the outcome conditional on both provider and covariates; rather, it is sufficient to estimate the outcome conditional only on covariates. This greatly simplifies the estimation problem, as it obviates the need to estimate provider-covariate interactions. However, this simplification only occurs under the particular choice of setting $P_{Z|W} = P_{A|W}$. 

The required identification assumptions for the direct and indirect parameters foreshadows their relative difficulty for estimation. The direct standardization parameter requires a strong positivity assumption because it must be possible to estimate what would have happened if every member of the population had been treated by a particular provider. Such an assumption may not hold in the real world for situations where there are many providers or there are providers who never or rarely treat a subpopulation of patients. In contrast, the stochastic intervention underlying the indirect standardization parameter is designed such that patients are counterfactually randomized to providers who already had a positive probability of having treated them, thereby avoiding the need for a strong positivity assumption. There is a caveat, however. The direct parameters are standardized based on a common reference population, they can be used to directly compare providers against one another. The indirect parameters, however, cannot be used to compare providers directly as each indirectly standardized outcome is based on a different reference population \citep{shahian2008, shahian2020misuse}.


\section{Efficiency Theory}
\label{section:efficiency-theory}
In this section we study the semiparametric properties of the standardization parameters. Our analysis is based on establishing the \textit{efficient influence function} (EIF) of the target statistical parameters \citep{ vanderVaart98}. The EIF of a parameter is an important object of interest in nonparametric efficiency theory, as its variance defines the efficiency bound for estimating the parameter in a nonparametric model \citep{Bickel97}. In addition, using knowledge of the form of the EIF we can define estimators that achieve the efficiency bound \citep{vanderLaanRose11}. Estimators of a parameter $\psi : \mathcal{M} \to \mathbb{R}$ based on the EIF $\D(P) : \mathcal{O} \to \mathbb{R}$ of $\psi$ at $P \in \mathcal{M}$ derive good statistical properties based on the following von-Mises expansion \citep{mises1947asymptotic}:
\begin{align}
    \psi(P) = \psi(P_0) - \E_{P_0}\left[ \D(P)(O) \right] + \R(P, P_0),
\end{align}
where $\R$ is a second-order remainder term. Applying the expansion with an estimate $\hat{P}$ of $P_0$ shows the bias of an arbitrary plug-in estimator can be approximated by the expectation of the EIF. One strategy for forming an unbiased estimator is therefore to simply add to the initial estimate the empirical mean of the estimated influence function; this serves as the basis of so-called one-step estimation \citep{pfanzagl1982,emery2000} and debiased/double machine learning \citep{chernozhukov2018dml}. Alternatively, one could carefully fluctuate the initial estimate $\hat{P}$ to form an updated estimate $\hat{P}^*$ designed to have the property that the empirical mean of the EIF evaluated at $\hat{P}^*$ is zero; this is the approach taken by targeted minimum loss-based estimation \citep{vanderLaan2003unifiedmethods, vanderLaanRose11}. For detailed introductions to semiparametric efficiency theory with applications to causal inference, excellent overviews can be found in \cite{kennedy2016efficiency} and \cite{kennedy2023semiparametric}. 

Theorem~\ref{theorem:eif-direct} restates the EIF of the direct standardization parameter $\phi(P_0)(a')$. This result is well-known, as direct standardization is equivalent to the mean counterfactual outcome under a categorical exposure, which has been previously studied.
\begin{theorem}[Efficient influence function for direct parameter]
\label{theorem:eif-direct}
The efficient influence function $\D_{a'}(P)$ of the parameter $\direct(P)(a')$ at $P$ is given by
\begin{align}
    \D_{a'}(P)(o) = \frac{\mathbb{I}[a = a']}{\g_P(a', w)} \left\{ y - \Qaw_P(a', w) \right\} + \Qaw_P(a', w) - \phi(P)(a').
\end{align}
\end{theorem}
The proof of all theorems can be found in the Supplemental Material. Theorem~\ref{theorem:eif-psi1} gives the EIFs of $\psi_1$, $\psi_2$, $\psi_{\excessrisk}$ and $\psi_{\smr}$. To the best of our knowledge, this is the first time the EIFs of the indirect standardization parameters $\psi_{\excessrisk}$ and $\psi_{\smr}$ have been presented. 
\begin{theorem}[Efficient influence functions for indirect parameters]
\label{theorem:eif-psi1}
For all $a' \in \{1, \dots, m\}$, the parameters $\psi_1(P)(a')$, $\psi_2(P)(a')$, $\psi_{\excessrisk}(P)(a')$ and $\psi_{\smr}(P)(a')$  have EIFs at $P \in \mathcal{M}$ given by
\begin{align}
    \D_{2,a'}(P)(o) &= \frac{1}{P(A = a')} \Bigg\{ \g_P(a', w) \left( y - \Qw_P(w) \right) + \mathbb{I}[a = a'] \left(\Qw_P(w) - \psi_2(P)(a')\right)\Bigg\}, \\
    \D_{\excessrisk, a'}(P)(o) &= \D_{1,a'}(P)(o) - \D_{2,a'}(P)(o), \\
    \D_{\smr, a'}(P)(o) &= \frac{1}{\psi_2(P)(a')} \D_{1,a'}(P)(o) - \frac{\psi_1(P)(a')}{\psi_2(P)(a')^2} \D_{2,a'}(P)(o).
\end{align}
\end{theorem}
The form of the EIFs $\D_{a'}(P)$ for the direct parameter $\direct$ and and $\D_{2,a'}(P)$ indirect parameter $\indirect_2$ provide important insight into the statistical properties of the two parameters. Notably, the EIF of the direct parameter is a function of the \textit{inverse} of the probability of treatment $\g_P(a', w)$. Therefore, the existence of subpopulations who are unlikely to be treated by provider $a'$ cause the efficiency bound for estimating the associated direct standardization parameter to increase. This reflects why in many real-world applications directly standardized outcomes are difficult to estimate when case-mixes vary significantly between providers. On the other hand, the EIF for the indirect parameter $\indirect_2$ depends directly on the probability of treatment $\g_P(a', w)$, rather than on its inverse. Strata with low probability of being treated by provider $a'$ therefore \textit{decrease} the efficiency bound for estimating the indirectly standardized outcome parameter. As a result, the indirect standardization parameter will typically be easier to estimate than the direct standardization parameter when there are strata of covariates for which $\g_P(a', w)$ is small. 

Establishing the form of the second-order remainder term $R(P, P_0)$ of the von-Mises expansion is important for finding conditions under which $R(\hat{P}, P_0)$ converges to zero sufficiently quickly. The parameters $\phi$, $\psi_1$ and $\psi_2$ admit von-Mises expansions with second-order remainder terms given below.
\begin{theorem}[von-Mises expansions]
    \label{theorem:von-mises}
    The parameters $\phi(P)(a')$, $\psi_{1}(P)(a')$ and $\psi_{2}(P)(a')$,  admit the representations
    \begin{align}
        \phi(P)(a') &= \phi(P_0)(a') - \E_{P_0}[\D_{a'}(P)(O)] + \R_{a'}(P, P_0), \\
        \psi_1(P)(a') &= \psi_1(P_0)(a') - \E_{P_0}[\D_{1,a'}(P)(O)] + \R_{1,a'}(P, P_0), \\
        \psi_2(P)(a') &= \psi_1(P_0)(a') - \E_{P_0}[\D_{2,a'}(P)(O)] + \R_{2,a'}(P, P_0),
    \end{align}
    where the second-order remainder terms are given by
    \begin{align}
        \R_{a'}(P, P_0) &= \E_{P_0}\left[\left(\g_P(a', W) - \g_{0}(a', W) \right) \left( \Qw_{0}(W) - \Qw_P(W) \right) \right], \\
        \R_{1,a'}(P, P_0) &= \E_{P_0}\left[ \left( \frac{P_0(A = a')}{P(A = a')} - 1 \right) \left( \Qa_P(A) - \Qa_{0}(A) \right) \right], \\
        \R_{2, a'}(P, P_0) &= \E_{P_0}\left[ \frac{1}{P(A = a')} \left(\g_P(a', W) - \g_{0}(a', W) \right) \left( \Qw_{0}(W) - \Qw_P(W) \right) \right].
    \end{align}
\end{theorem}
The form of the second-order remainder terms $\R_{a'}$ and $\R_{2, a'}(P, P_0)$ suggests the possibility of constructing doubly robust estimators of $\phi(P_0)(a')$ and $\psi_2(P_0)(a')$ (and, subsequently, of $\psi_\excessrisk(P_0)(a')$ and $\psi_{\smr}(P_0)(a')$), in that as long as either $g_{0}$ or $\Qw_0$ is estimated consistently, the overall remainder term will go to zero.

\section{Estimation}
\label{section:estimation}
Now that we have established the semi-parametric efficiency bounds for estimating the direct and indirect standardization parameters, we turn to the task of finding practical estimators that achieve these bounds. To begin,
suppose we have initial estimators of the parts of $P$ relevant to the target parameters and their EIFs. For both direct and indirect standardization parameters, we require an estimate of the empirical distribution of the covariates $\lambda(W)$, for which we will use the empirical distribution. The propensity score model can be estimated using logistic regression or any other equivalent algorithm for binary or multinomial regression (for example, \cite{daignault2017doublerobust} use multinomial logistic regression), which may be data-adaptive. The mean outcome for each provider $\Qa$ can be estimated using simple empirical means. The outcome models $\Qaw_0$ and $\Qw_0$ can also be estimated using any regression method, such as generalized linear models or via data-adaptive machine learning algorithms. Notationally, we will write $\theta_n^0$ to denote an initial estimate of a variable $\theta_0$ (for example, $\Qw_n^0$ denotes an initial estimate of $\Qw_0$). 

We rely on cross-fitting in order to avoid complexity conditions on the class of algorithms used to estimate $\g_0$, $\Qaw_0$, and $\Qw_0$ \citep{zheng2011cvtmle}. Specifically, let $\mathcal{V}_1, \dots, \mathcal{V}_J$ be a disjoint partition of the observation indices $\{1, \dots, n \}$ into sets of validation indices, formed such that sets are of roughly equal size. Let $\mathcal{T}_1, \dots, \mathcal{T}_J$ be the associated training sets, with $\mathcal{T}_j = \{ 1, \dots, n \} \backslash \mathcal{V}_j$. Let $\theta_{n_j}^0$ denote the initial estimate of a generic parameter $\theta_0$ estimated using only the data corresponding to the training set $\mathcal{T}_j$. Let $j[i]$ be the training set corresponding to observation $i$. 


\subsection{Targeted Minimum Loss-Based Estimation}
Targeted minimum loss-based estimation (TMLE) is a framework for constructing asymptotically efficient plugin estimators of parameters in semiparametric models \citep{vanderLaanRose11}. TMLE for a parameter $\psi$ works by carefully fluctuating an initial estimate of $P$ in order to solve the empirical EIF estimating equation. The first step is to define, for any $P \in \mathcal{M}$, a parametric submodel $\{ P_\epsilon \}$ indexed by a finite-dimensional parameter $\epsilon$ and a loss function $\mathcal{L}$.
The submodel and loss function must be chosen carefully to have to two key properties: when $\epsilon = 0$, $P_\epsilon = P$ (that is, $P_\epsilon$ fluctuates $P$) and the EIF of $\psi$ at $P$ must be included in the linear span of the gradient of $\mathcal{L}(P_\epsilon)$ evaluated at $\epsilon = 0$. The parameter $\epsilon$ is subsequently estimated by minimizing the loss function $\mathcal{L}$ under the submodel $\{ P_\epsilon \}$. The gradient of the loss function evaluated at the minimizer $\hat{\epsilon}$ will be zero, and by construction of the submodel this implies the empirical EIF will also be zero. This takes care of the bias term in the von-Mises expansion. Via cross-fitting we can show that the second-order remainder terms also converge to zero. Taken together, these properties imply that the estimators will be unbiased.

The TMLE estimators for $\phi$, $\psi_1$, and $\psi_2$ have a similar structure. For the fluctuation models, for any distribution $P \in \mathcal{M}$ define the submodels $\{ \Qaw^\epsilon_P : \epsilon \in \mathbb{R} \}$, $\{ \Qa^\epsilon_P : \epsilon \in \mathbb{R} \}$, and $\{ \Qw^\epsilon_P : \epsilon \in \mathbb{R} \}$ as
\begin{align}
    \logit(\Qaw_P^\epsilon(A, W)) &= \logit(\Qaw_P)(A, W) + \epsilon M_P(A, W), \\
    \logit(\Qa_P^\epsilon(A)) &= \logit(\Qa_P)(A) + \epsilon K_P(A), \\
    \logit(\Qw_P^\epsilon(W)) &= \logit(\Qw_P)(W) + \epsilon H_P(W),
\end{align}
where
\begin{align}
    M_P(A, W) = \frac{\mathbb{I}[A = a']}{\g_P(a', W)}, \qquad K_P(A) = \frac{\mathbb{I}[A = a']}{P(A = a')}, \qquad H_P(W) = \frac{\g_P(a', W)}{P(A = a')}.
\end{align}
Note that for $\epsilon = 0$, then $\Qaw^\epsilon_P = \Qaw_P$, $\Qa^\epsilon_P = \Qa_P$, and $\Qw^\epsilon_P = \Qw_P$. The parameter $\epsilon \in \mathbb{R}$ is estimated independently in each submodel via maximum likelihood estimation with corresponding loss functions
\begin{align}
    \bar{\mathcal{L}}(\epsilon) &= \sum_{i=1}^N A_i \log \Qaw_P^\epsilon(A_i, W_i) + (1 - A_i) \log(1 - \Qaw_P^\epsilon(A_i, W_i)), \\
    \dot{\mathcal{L}}(\epsilon) &= \sum_{i=1}^N A_i \log \Qa_P^\epsilon(A_i) + (1 - A_i) \log(1 - \Qa_P^\epsilon(A_i)), \\
    \tilde{\mathcal{L}}(\epsilon) &= \sum_{i=1}^N A_i \log \Qw_P^\epsilon(W_i) + (1 - A_i) \log(1 - \Qw_P^\epsilon(W_i)).
\end{align} 
The initial estimates are then updated:
\begin{align}
    \Qaw_P^*(A, W) &= \logit^{-1}(\logit(\Qaw^0_P(A, W) + \hat{\epsilon} M_P(A, W)), \\
    \Qa_P^*(A) &= \logit^{-1}(\logit(\Qa^0_P(A) + \hat{\epsilon} K_P(A)), \\
    \Qw_P^*(W) &= \logit^{-1}(\logit(\Qw^0_P(W) + \hat{\epsilon} H_P(W)),
\end{align}
where $\hat{\epsilon}$ is the MLE of $\epsilon$ under the corresponding loss function. 
The TMLE estimates $\phi^{\tmle}$, $\psi_1^\tmle$, and $\psi_2^\tmle$ then be formed by plugging in the updated parameters $\Qaw_P^*$, $\Qa_P^*$, and $\Qw_P^*$ into the parameter mappings. The excess risk and standardized event ratios can be estimated by plugging in the TMLE estimates of $\psi_1$ and $\psi_2$:
\begin{align}
    \psi^\tmle_\excessrisk = \psi_1^\tmle - \psi_2^\tmle, \qquad \text{and} \qquad \psi^\tmle_{\smr} = \frac{\psi^\tmle_1}{\psi^\tmle_2}. 
\end{align}
In practice, the fluctuations are based on the cross-fitted estimates of all the required nuisance parameters.
Under conditions, the TMLE estimators are asymptotically normal and efficient. 
\begin{theorem}[Asymptotic normality and efficiency of TMLE]
\label{theorem:tmle-normality}
Fix $a' \in \{ 1, \dots, m \}$. 
\begin{enumerate}
    \item Assume that $\R_{a'}(P, P_0) = o_P(n^{-1/2})$. Then
    \begin{align}
        \sqrt{n}(\phi^{\tmle}(a') - \phi(P_0)(a')) \leadsto N(0, P_0[\D_{a'}(P_0)^2]).
    \end{align}
    \item Assume that $\R_{1,a'}(P, P_0) = o_P(n^{-1/2})$. Then
    \begin{align}
        \sqrt{n}(\psi_1^{\tmle}(a') - \psi_1(P_0)(a')) \leadsto N(0, P_0[\D_{1,a'}(P_0)^2]).
    \end{align}
    \item Assume that $\R_{2, a'}(P, P_0) = o_P(n^{-1/2})$. Then
    \begin{align}
        \sqrt{n}(\psi_2^{\tmle}(a') - \psi_2(P_0)(a')) \leadsto N(0, P_0[\D_{2,a'}(P_0)^2]).
    \end{align}
    \item Assume that $\R_{1, a'}(P, P_0) = o_P(n^{-1/2})$ and $\R_{2, a'}(P, P_0) = o_P(n^{-1/2})$. Then 
    \begin{align}
        \sqrt{n}(\psi_{\excessrisk}^{\tmle}(a') - \psi_{\smr}(P_0)(a')) \leadsto N(0, P_0[\D_{\excessrisk,a'}(P_0)^2]), \\
        \sqrt{n}(\psi_{\smr}^{\tmle}(a') - \psi_{\smr}(P_0)(a')) \leadsto N(0, P_0[\D_{\smr,a'}(P_0)^2]).
    \end{align}
\end{enumerate}
\end{theorem}

The assumption that $\R_{1,a'}(P, P_0) = o_P(n^{-1/2})$ is easily satisfied if $P(A = a')$ is estimated using the estimator $P_n \mathbb{I}[a_i = a]$ (or, more precisely, the cross-fitted equivalent) by the law of large numbers. The assumptions that $\R_{a'}(P, P_0) = o_p(1)$ and $\R_{2,a'}(P, P_0) = o_p(1)$ are satisfied if $\g_0$, $\Qaw_0$, and $\Qw_0$ are estimated using correctly specified parametric models, as in that case each would converge at parametric $n^{-1/2}$ rates. However, due to the product structure of the remainder terms $\R_{a'}$ and $\R_{2,a'}$, the assumption also holds as long as the estimators converge faster than $n^{-1/4}$. 

We note that the above asymptotic normality results also extend to joint distributions of the parameters across different providers. For example, the set $\{ \psi^{\tmle}(a') : a' \in \{1, \dots, m \} \}$ converges to a multivariate normal distribution with covariance matrix characterized for row $a_1$ and column $a_2$ by $P_0[\D_{a_1}(P_0)(O) \D_{a_2}(O)]$. One could use this fact to simulate a reference distribution of providers against which individual providers could be compared. For example, the median provider direct standardization parameter could be computed and used as a reference against which each provider's direct standardization parameter is compared.

The product structure of the second-order remainder terms implies that the TMLE estimators may be consistent even if one of the two nuisance parameters is not consistent. This property is often referred to as \textit{double robustness}, which we formalize in the following theorem.
\begin{theorem}[Double robustness]
    \label{theorem:double-robustness}
    \leavevmode
    \begin{enumerate}
        \item Suppose that either $\| \g_P - \g_0 \|_{L_2(P_0)} = o_p(1)$ or $ \| \Qaw_P - \Qaw_0 \|_{L_2(P_0)} = o_p(1)$. Then
        \begin{align}
            \phi^{\tmle}(a') - \phi(P_0)(a') &= o_p(1).
        \end{align}
        \item Suppose that either $\| \g_P - \g_0 \|_{L_2(P_0)} = o_p(1)$ or $\| \Qw_P - \Qw_0 \|_{L_2(P_0)} = o_p(1)$. Then
        \begin{align}
            \psi_2^{\tmle}(a') - \psi_2(P_0)(a') &= o_p(1), \\ 
            \psi^\tmle_\excessrisk(a') - \psi_{\excessrisk}(P_0)(a') &= o_p(1),\\
            \psi_{\smr}^{\tmle}(a') - \psi_{\smr}(P_0)(a') &= o_p(1). 
        \end{align}
    \end{enumerate}
\end{theorem}
The proof of Theorem~\ref{theorem:double-robustness} follows straightforwardly from the form of the second-order remainder terms of the von-Mises expansions of the parameters and is therefore omitted. 

\section{Simulation Study}
\label{section:simulations}
In the previous section, we established asymptotic guarantees for the proposed targeted estimators. In this section, we investigate finite sample performance via  simulation studies. The first simulation study is designed to illustrate the potential benefits of flexible data-adaptive algorithms for estimating nuisance parameters rather than using generalized linear models. To that end, we focus on a rather simple data generating process. Our goal is to show that even in simple situations, generalized linear models can yield biased estimates of direct and indirect standardization parameters. In the second simulation study we expand to a more complex data generating process to study the performance of the proposed estimators when there are a larger number of covariates and providers. In the interest of succinctness, we focus in the simulation studies on data structures with continuous outcomes.

The two simulation studies can be described in a common framework. Let $a \in \{ 1, \dots, m \}$ index providers. For each provider, we drew a provider effect following $\beta_a \sim \mathrm{Bernoulli}(0.5)$. Each patient has a set of covariates drawn from $(W_1, \dots, W_k) \sim \mathrm{Unif}^k(0, 1)$, where $k$ is a positive integer and $\mathrm{Unif}^k$ denotes the $k$-dimensional joint uniform distribution. The patient is assigned to provider $a \in \{1, \dots, m \}$ by a multinomial draw with probability 
\begin{align}
    P(A = a | W) \propto g(\beta_a, W),
\end{align}
where $g$ is a fixed function that is specified differently for each of the simulation studies. The function $g$ is normalized such that $\sum_{a=1}^m \g(a, W) = 1$. The patient's outcome is then drawn according to
\begin{align}
Y | A, W \sim \mathrm{N}\left( Q(W, A, \beta_A), \sigma^2 \right),
\end{align}
where $\sigma > 0$ is fixed and $Q$ is a function specified in each simulation study.

To summarize overall performance of each estimator, the mean error (ME), mean  absolute error (MAE), and 95\% empirical coverage of $\phi^{\tmle}(a')$, $\psi_1^{\tmle}(a')$, $\psi_2^{\tmle}(a')$, $\psi_{\excessrisk}^{\tmle}(a')$ and $\psi_{\smr}^{\tmle}(a')(a')$ were calculated for each $a' \in \{1, \dots, m \}$ and subsequently averaged.

The simulation studies were conducted using the \texttt{R} statistical computing environment \citep{r2022} and the \texttt{TargetedRisk} package, available at \url{https://github.com/herbps10/TargetedRisk}. Materials for reproducing the simulation studies is available at \url{https://github.com/herbps10/ProviderProfiling}. 

\subsection{Simulation Study 1}
The first simulation study is designed with the narrow goal of comparing TMLE to a more traditional GLM estimator in a scenario where the propensity score and outcome models are nonlinear. The data generating process is simple. Each patient has a single covariate ($k = 1$). The provider assignment probabilities are defined by
\begin{align}
    \mathrm{logit}^{-1}(g(\beta_A, W)) = \begin{cases}
        2, & W_1 > 0.7, \beta_A = 1, \\
        -2, & W_1 \leq 0.7, \beta_A = 1, \\
        -2, & W_1 > 0.7, \beta_A = 0, \\
        2, & W_1 \leq 0.7, \beta_A = 0.
    \end{cases}
\end{align}
That is, patients with $W_1 > 0.7$ are more likely to be assigned to hospitals with hospital effect $\beta_A = 1$. Such a relationship may exist in real-world scenarios if, for example, $W_1$ reflects a measure of disease severity and patients with higher disease severity are preferentially assigned to hospitals that specialize in that disease. The outcome function was then defined via the piecewise function
\begin{align}
    Q(W, A, \beta_A) = \begin{cases}
        0.3, & \beta_A = 1, 0 \leq W_1 \leq 0.5 \\
        1,   & \beta_A = 1, 0.5 < W_1 \leq 0.7 \\
        2,   & \beta_A = 1, 0.7 < W_1 \leq 1, \\
        0.7, & \beta_A = 0, 0 \leq W_1 \leq 0.5, \\
        0.5, & \beta_A = 0, 0.5 < W_1 \leq 0.7, \\
        0    & \beta_A = 0, 0.7 < W_1 \leq 1. \\
    \end{cases}
\end{align}

Next, $100$ simulation datasets were created each having $m = 10$ providers and with sample sizes $N = \{ 1000, 2500, 5000 \}$ total patients. As a benchmark, generalized linear models were used to estimate the direct and indirect parameters (see Supplemental Material for details). The proposed TMLE estimator was applied with nuisance parameters estimated using Super Learning. The Super Learning library included two random forest configurations (\texttt{ranger} with $50$ and $100$ trees and with the max tree depth set to $2$).  

The results for the parameters $\direct$, $\indirect_{2}$, $\indirect_{\smr}$, and $\indirect_{\excessrisk}$  are shown in Table~\ref{tab:simulation_study_1}. Results for $\indirect_{1}$ are not shown as it is a simple mean of observed variables and not a true counterfactual quantity. The mean error of the TMLE estimator decreases with sample size for all parameters, while the same is not true of the GLM estimator. The coverage of the 95\% confidence intervals derived from the TMLE estimators was near nominal for the indirect parameters and slightly conservative for the direct parameter.  

\begin{table}[hbtp]
    \centering
    \begin{tabular}{lrrrrr}
    \hline
    & \multicolumn{2}{c}{ME} & \multicolumn{2}{c}{MAE} & 95\% Coverage \\
    $N$ & TMLE & GLM & TMLE & GLM & TMLE \\
    \hline
    \multicolumn{6}{l}{$\direct$} \\
    1000 & 0.541 & 0.100 & 0.575 & 0.107 & 99.2\%\\
    2500 & 0.001 & 0.102 & 0.027 & 0.104 & 99.6\%\\
    5000 & 0.000 & 0.100 & 0.017 & 0.101 & 98.9\%\\
    \multicolumn{6}{l}{$\indirect_{2}$} \\
    1000 & -0.007 & 0.031 & 0.026 & 0.108 & 91.6\%\\
    2500 & -0.002 & 0.031 & 0.015 & 0.109 & 92.6\%\\
    5000 & -0.002 & 0.028 & 0.011 & 0.107 & 93.7\%\\
    \multicolumn{6}{l}{$\indirect_{\smr}$} \\
    1000 & -0.002 & -0.066 & 0.044 & 0.185 & 93.7\%\\
    2500 & -0.001 & -0.064 & 0.028 & 0.186 & 94.0\%\\
    5000 & 0.000 & -0.062 & 0.019 & 0.183 & 94.6\%\\
    \multicolumn{6}{l}{$\indirect_{\excessrisk}$} \\
    1000 & -0.002 & -0.030 & 0.029 & 0.108 & 92.9\%\\
    2500 & -0.001 & -0.030 & 0.018 & 0.109 & 93.6\%\\
    5000 & 0.000 & -0.028 & 0.012 & 0.108 & 94.1\%\\
    \hline
    \end{tabular}
    \caption{Results from Simulation Study 1 showing mean error (ME) and mean absolute error (MAE) for the TMLE and GLM estimators. The empirical coverage of the 95\% confidence intervals derived from the TMLE estimator are also shown. }
    \label{tab:simulation_study_1}
\end{table}

\subsection{Simulation Study 2}
Simulation study 2 has a larger scope than simulation study 1, and investigates the performance of the TMLE estimators in a more complex setting. We also probe the doubly robust properties of the TMLE estimators when nuisance parameters are estimated inconsistently. In these simulations, each patient has $k = 10$ covariates. The provider assignment mechanism is defined by
\begin{align}
    g(\beta_A, W)) \propto 1 + 10 \beta_A W_1 - 0.2 W_2 - 0.5 W_5,
\end{align}
suitably normalized such that $\sum_{a} g(\beta_a, W) = 0$. The outcome model is given by
\begin{align}
    Q(W, A, \beta_A) = 2 W_1 - (W_2 - 0.5)^2 + \mathbb{I}[W_3 > 0.5] + W_5 - 2\beta_A.
\end{align}
The nuisance parameters are estimated in four configurations:
\begin{itemize}
    \item Scenario 1: $\Qw_0$ is estimated using SuperLearner with \texttt{mean}, \texttt{glm}, \texttt{glmnet}, \texttt{ranger} and \texttt{knn} learners and $\g_0$ estimated using SuperLearner with \texttt{mean}, \texttt{glm}, \texttt{glmnet}, and \texttt{ranger} learners.
    \item Scenario 2: $\Qw_0$ estimated as in Scenario 1 and $\g_0$ misspecified.
    \item Scenario 3: $\g_0$ estimated as in Scenario 1 and $\Qw_0$ misspecified.
    \item Scenario 4: $\Qw_0$ and $\g_0$ both misspecified.
\end{itemize}
To ensure model misspecification, we supplied the TMLE method with $\Qw_0$ estimates fixed to $0.5$ and randomly generated estimates of $\g_0$ (following a Dirichlet distribution). As such, scenarios 1-3 represent a type of ``worst-case" model misspecification. The simulation data generating process was sampled from $100$ times with each dataset comprising $m = 75$ providers and $N \in \{ 5\times10^3, 1\times10^4, 5\times10^4 \}$ patients.

Simulation results are shown in Table~\ref{tab:simulation_results} for the directly standardized parameter $\direct$ and in Table~\ref{tab:simulation_results2} for the indirectly standardized parameters $\indirect_2$, $\indirect_{\excessrisk}$ and $\indirect_{\smr}$. In Scenario 1 when both $\Qw_0$ and $g_0$ are estimated using flexible ensembles the TMLE estimators exhibit decreasing mean and mean absolute error with increasing sample size and have near optimal empirical coverage. On the other hand, in Scenario 4 where both $\Qw_0$ and $g_0$ are estimated using misspecified models, the estimators TMLE estimators exhibit significant bias and do not have correct empirical coverage. In Scenarios 2 and 3, when either $\Qw_0$ or $g_0$ are not estimated correctly, the estimators again have errors that decrease as sample size increases, which is to be expected from the theoretical doubly robust properties of the TMLE estimators. In comparison, the GLM estimator is biased at all sample sizes. 

\begin{table}[hbtp]
    \centering
    \begin{tabular}{lrrrr}
        \hline
        & & \multicolumn{3}{c}{$\direct$} \\
        Estimator & $N$ & ME & MAE & Coverage \\
        \hline
        GLM              & 5,000  & 0.148 & 0.439 & $-$ \\
                         & 10,000 & 0.151 & 0.418 & $-$ \\
                         & 20,000 & 0.155 & 0.413 & $-$ \\
        TMLE, Scenario 1 & 5,000 & -0.368  & 0.568 & 95.2\%\\
                         & 10,000 & -0.021 & 0.137 & 97.3\%\\
                         & 20,000 & -0.020 & 0.059 & 97.2\%\\
        TMLE, Scenario 2 & 5,000 & -0.362  & 0.368 & 48.4\%\\
                         & 10,000 & -0.252 & 0.257 & 49.8\%\\
                         & 20,000 & -0.203 & 0.207 & 49.8\%\\
        TMLE, Scenario 3 & 5,000 & 0.374  & 0.837 & 95.3\%\\
                         & 10,000 & 0.386 & 0.611 & 96.6\%\\
                         & 20,000 & 0.356 & 0.547 & 89.7\%\\
        TMLE, Scenario 4 & 5,000 & 0.769  & 0.822 & 49.5\%\\
                         & 10,000 & 0.790 & 0.813 & 50.2\%\\
                         & 20,000 & 0.801 & 0.808 & 49.9\%\\
        \hline
    \end{tabular}
    \caption{Results from Simulation Study 1 for the direct standardization parameter $\direct$ showing mean error (ME), mean absolute error (MAE), and 95\% empirical coverage for the GLM estimator and for TMLE under nuisance parameter misspecification scenarios.}
    \label{tab:simulation_results}
\end{table}

\begin{table}[hbtp]
    \centering
    \resizebox{\columnwidth}{!}{
    \begin{tabular}{lrrrrrrrrrr}
        \hline
        & & \multicolumn{3}{c}{$\psi_2$} & \multicolumn{3}{c}{$\indirect_\excessrisk$ (difference)} & \multicolumn{3}{c}{$\indirect_{\smr}$ (ratio)} \\
        Estimator & $N$ & ME & MAE & Coverage & ME & MAE & Coverage & ME & MAE & Coverage \\
        GLM & 5,000 & -0.220 & 0.409 & $-$ & 0.217 & 0.403 & $-$ & 0.287 & 0.337 & $-$ \\
        & 10,000 & -0.213 & 0.389 & $-$ & 0.214 & 0.392 & $-$ & 0.296 & 0.343 & $-$\\
        & 20,000 & -0.213 & 0.388 & $-$ & 0.215 & 0.387 & $-$ & 0.295 & 0.341 & $-$ \\
        TMLE, Scenario 1 & 5,000 & -0.055 & 0.212 & 92.8\% & -0.010 & 0.102 & 91.7\% & -0.040 & 0.163 & 92.2\%\\
        & 10,000 & -0.025 & 0.145 & 93.5\% & -0.004 & 0.064 & 93.1\% & -0.019 & 0.108 & 93.7\%\\
        & 20,000 & -0.012 & 0.099 & 94.5\% & -0.001 & 0.043 & 94.6\% & -0.005 & 0.071 & 94.5\%\\
        TMLE, Scenario 2 & 5,000 & -0.057 & 0.210 & 93.0\% & -0.007 & 0.101 & 92.0\% & -0.031 & 0.158 & 92.1\%\\
        & 10,000 & -0.027 & 0.144 & 93.6\% & -0.002 & 0.064 & 93.3\% & -0.012 & 0.106 & 93.4\%\\
        & 20,000 & -0.014 & 0.099 & 94.7\% & 0.001 & 0.043 & 94.6\% & -0.001 & 0.070 & 94.3\%\\
        TMLE, Scenario 3 & 5,000 & 0.299 & 0.446 & 36.8\% & -0.363 & 0.455 & 61.1\% & -0.108 & 0.252 & 59.2\%\\
        & 10,000 & 0.114 & 0.323 & 22.7\% & -0.143 & 0.320 & 53.5\% & 0.052 & 0.206 & 45.6\%\\
        & 20,000 & 0.003 & 0.291 & 7.1\% & -0.016 & 0.286 & 27.9\% & 0.136 & 0.222 & 23.0\%\\
        TMLE, Scenario 4 & 5,000 & -0.444 & 1.441 & 0.0\% & 0.379 & 1.419 & 3.2\% & 0.357 & 0.717 & 4.7\%\\
        & 10,000 & -0.354 & 1.386 & 0.0\% & 0.325 & 1.382 & 0.1\% & 0.355 & 0.724 & 0.2\%\\
        & 20,000 & -0.321 & 1.371 & 0.0\% & 0.308 & 1.367 & 0.0\% & 0.348 & 0.719 & 0.0\%\\
        \hline
    \end{tabular}
    }
    \caption{Results from Simulation Study 2 for the indirect standardization parameters $\indirect_2$, $\indirect_{\excessrisk}$, and $\indirect_{\smr}$ showing mean error (ME), mean absolute error (MAE), and 95\% empirical coverage of the TMLE estimators.}
    \label{tab:simulation_results2}
\end{table}

\section{Application}
\label{section:application}
We illustrate our methods by analyzing a Medicare claims dataset for dialysis providers treating patients with end-stage renal disease (ESRD) supplied by the United States Renal Data System (USRDS, \citealt{usrds2022}). This dataset was previously analyzed in \citet{wu2024leveraging}, to which we refer to for a detailed description. In brief, the covariates $W$ include patient demographics, physical characteristics, social variables, clinical variables, and comorbidities (see the Supplemental Material for a full listing of covariates). The treatment variable $A \in \{1, \dots, m \}$ indexes the dialysis providers, identified by their USRDS assigned facility identifiers. The outcome variable $Y \in \{0, 1 \}$ indicates all-cause unplanned hospital readmission within 30 days of discharge ($Y = 1$ indicates an unplanned readmission). We focus on a subset of the dataset comprising providers in New York State having at least 100 observations each (a discussion of methods to handle evaluation of providers with few observations can be found in Section~\ref{section:discussion}). After subsetting, there are $m = 84$ dialysis providers and $N = 30,297$ observations in the analysis dataset.

We computed indirect standardization parameters for each of the providers in the analysis dataset using both the TMLE estimators and generalized linear models. In this context, the indirect ratio parameter $\indirect_{\smr}$ is also known as a standardized readmission ratio (SRR, see \citealt{he2013}). For the TMLE estimator, we estimated the nuisance parameters using ensembles that included \texttt{glmnet}, \texttt{glm}, and four versions of \texttt{lightgbm} with $100$, $250$, $500$, and $1000$ trees. The estimated propensity scores ranged from $1.6 \times 10^{-8}$ to $0.92$, suggesting extreme practical positivity violations in the dataset. As such, there is strong evidence that the positivity assumption~\ref{assumption:positivity} does not hold in this dataset. We therefore chose not to calculate directly standardized readmission rates. 

\begin{figure}
    \centering
    \includegraphics[width=0.7\linewidth]{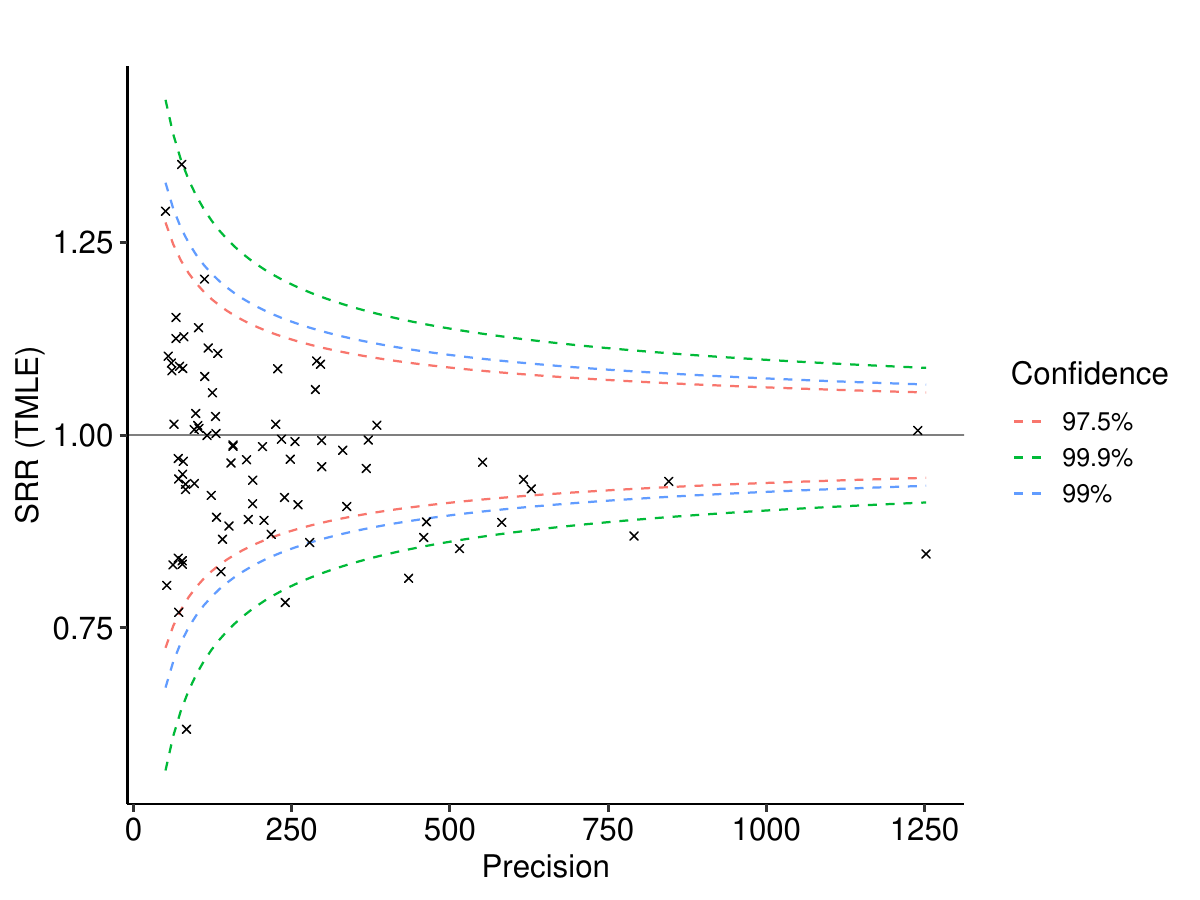}
    \caption{Funnel plot of standardized readmission ratios (SRR) estimated using TMLE for New York State dialysis providers in the analysis dataset. The dotted line shows the 95\%, 99\%, and 99.9\% confidence limits for the SRR being greater than or below unity.}
    \label{fig:srr-funnel}
\end{figure}

One way of visualizing the SRR results is via a funnel plot \citep{spiegelhalter2005funnel}, which plots the estimated SRR against the precision of the estimate. In Figure~\ref{fig:srr-funnel} we use the variance of the estimated EIF as the measure of precision for each estimated SRR. This definition of precision reflects the semi-parametric efficiency bound for estimating the SRR in a non-parametric model, as discussed previously. In addition, control limits can be easily plotted at varying levels of significance levels. The results indicate that there are several providers with an estimated SRR less than one at the $99.9\%$ level. No providers had an SRR higher than one at the 99.9\% level, although there was one that was higher than one at the 99\% level.

\begin{figure}
    \centering
    \includegraphics[width=0.5\linewidth]{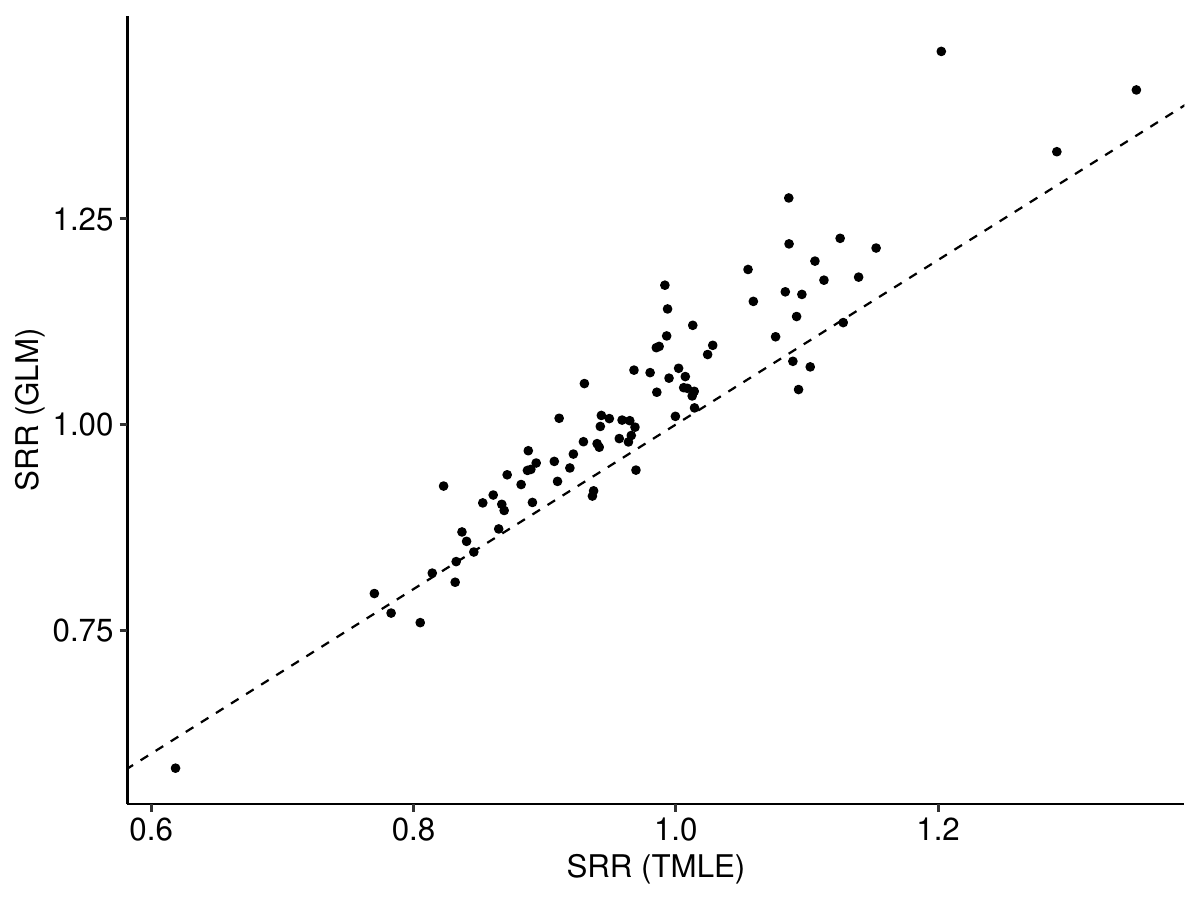}
    \caption{Comparison of Standardized Readmission Ratios estimated using GLM and with TMLE for New York State dialysis providers in the analysis dataset.}
    \label{fig:srr-comparison}
\end{figure}

The TMLE estimator yields different results than a GLM estimator of the SRR, as shown in Figure~\ref{fig:srr-comparison}. The GLM estimator tended to estimate higher SRR than the TMLE approach, which was more conservative. 

\section{Discussion}
\label{section:discussion}
In this work we reviewed the definition of direct and indirect outcome standardization parameters within a causal inference framework, and contributed a novel analysis of the parameters in terms of their semiparametric efficiency properties. We then proposed targeted estimators that achieve the semiparametric efficiency bounds for the target parameters. Our methodology is an advance over previous approaches in that the TMLE estimator is formulated within a nonparametric framework. As a consequence, the nuisance parameters may be estimated using methods that converge at slower than $n^{-1/2}$ rates, which opens the door to the use of data-adaptive machine-learning algorithms or ensembles of algorithms. Such flexibility helps avoid biased estimation due to model misspecification.  

The results of the simulation studies illustrate the strengths of nonparametric doubly robust estimation for provider profiling. The first simulation study illustrates how, even in a very simple scenario, unwarranted linearity assumptions on the functional forms of the nuisance parameters may lead to biased profiling estimates. By contrast, using flexible machine learning algorithms to estimate nuisance parameters combined with TMLE yielded unbiased estimates with reasonable uncertainty quantification. The second simulation study shows that the targeted estimators perform well in a more complex setting with complex treatment and outcome assignment mechanisms. 

In our application, we investigated dialysis providers in terms of unplanned readmissions. We chose not to estimate directly standardized readmission rates due to evidence of extreme positivity violations, suggesting the presence of providers in this dataset who treat disparate patient populations. A benefit of our doubly robust approach is that such positivity violations can be readily identified, as the propensity scores must be calculated as a nuisance parameter. A possible extension of this approach could be to find groups of providers who treat similar populations, for example via clustering, such that directly standardized metrics can be well estimated within each group. 

We proceeded to use the proposed targeted estimators to calculate standardized readmission ratios for New York State dialysis providers. The resulting estimates identified several providers with an SRR less than one at the 99.9\% significance level, and one provider had an SRR greater than one at the 99\% significance level.  A causal interpretation of these findings would be that patients treated by these providers fared better in terms of readmission than if they had been randomly assigned to a similar provider (in terms of case-mix). However, attributing such a causal interpretation to these findings is only valid insofar as the causal identifying assumptions are thought to hold. For the SRR, the key assumption is that of no unmeasured confounding (Assumption \ref{assumption:exchangeability}). While in this application we were able to include a rich set of possible confounding variables, there may remain additional unmeasured confounders. Therefore, we cannot assign a strictly causal interpretation to these results. A natural direction for future research is to develop sensitivity analyses to better understand the robustness of the estimated SRRs to the presence of unmeasured confounders. 


So far, provider profiling has focused primarily outcomes measured at a single time point. However, in many situations it may also be of interest to evaluate providers using longitudinal data measuring patient outcomes over time. To this end, it's interesting to note that 
the indirect standardization parameter $\psi_2(P)(a')$ can be interpreted as a type of Longitudinal Modified Treatment Policy (LMTP) with a stochastic intervention and conditioned on the treatment variable \citep{diaz2023lmtp}. This interpretation opens the door to defining and estimating provider  profiling parameters based on longitudinal data structures.

An important practical issue that often arises in provider profiling for small providers who see few patients. For these providers, finite sample effects may lead to high-variance estimates. As such, there is often interest in applying shrinkage estimators that partially pool provider quality estimates towards a common mean. A typical approach is to use generalized linear models with provider random effects that are shrunk towards zero. One way to address the issue using our estimators could be to apply Bayesian TMLE methods \citep{diaz2011} combined with Bayesian hierarchical modeling to allow for profiling estimates from small providers to be partially pooled towards a common mean \citep{gelman2013bda}. 

\section*{Acknowledgments}
We gratefully acknowledge support from the Alzheimer's Association (AARG-23-1077773), National Institute on Aging (K02AG076883), and the Department of Population Health and Center for the Study of Asian American Health at the NYU Grossman School of Medicine (U54MD000538). The data reported here have been supplied by the United States Renal Data System (USRDS). The interpretation and reporting of these data are the responsibility of the author(s) and in no way should be seen as an official policy or interpretation of the U.S. government. The computational requirements for this work were supported in part by the NYU Langone High Performance Computing (HPC) Core's resources and personnel.



\bibliography{references}
\begin{appendices}
    \section*{Supplemental Material}
    \section{Identification of $\indirect_2(P)(a')$}
The following establishes the identification of the indirect standardization parameter $\indirect_2(P)(a')$: 
\begin{align}
    \psi_2(P)(a') &= \E_P[Y^Z | A = a'] \\
                  &= \E_P[\E_P[Y^Z | Z, A, W] | A = a'] & \text{(Law of total expectation)} \\
                  &= \E_P\left[\int \E_P[Y^Z | Z = z, A, W] P(Z = z | A, W) dz \middle| A = a'\right] \\
                  &= \E_P\left[\int \E_P[Y^Z | Z = z, W] P(Z = z | W) dz \middle| A = a'\right] & \text{(Because $Z \indep A | W$)} \\
                  &= \E_P\left[\int \E_P[Y_A | A = a, W] \g_P(a, W) da \middle| A = a'\right] & \text{(By definition of $Z$)} \\
                  &= \E_P\left[\int \E_P[Y | A = a, W] \g_P(a, W) da \middle| A = a'\right] & \text{(By Assumption~\ref{assumption:exchangeability})} \\
                  &= \E_P\left[\E_P[Y | W] | A = a'\right]. 
\end{align}

\section{Deriving Efficient Influence Functions}
We derive the EIF of the target parameters in two steps. First, we propose a putative EIF by assuming that all the observed variables are discrete. The key tool we then leverage is the functional delta method \citep[Appendix A.1]{vanderLaanRose11}. Once we have a putative EIF in hand, we prove that is is indeed the general EIF for the target parameter by showing that the remainder term of the von-Mises expansion is second-order. 

\subsection{Proof of Theorem~\ref{theorem:eif-psi1}}
Fix $a' \in \{1, \dots, m\}$. We go into the most detail in deriving the EIF of $\psi_1(a')$, as the subsequent proofs are similar. We start by finding a putative EIF for $\psi_1$ by assuming the data are discrete and applying the functional delta method \citep[Appendix A.1]{vanderLaanRose11}. 

First, assume that $W$, $A$, and $Y$ are discrete. The nonparametric MLE of $\psi_1(P)(a')$ is given by
\begin{align}
    \hat{\psi}_1(P_n)(a') = \sum_{y \in \mathcal{Y}} y \frac{P_n f_{a',y}}{P_n f_{a'}},
\end{align}
where
\begin{align}
    f_{a',y}(O) &= \mathbb{I}[Y = y, A = a'], \\
    f_{a'}(O) &= \mathbb{I}[A = a'].
\end{align}
The derivatives of $\hat{\psi}_1(P)(a')$ with respect to each $f \in \mathcal{F}$ are given by
\begin{align}
    \frac{d\hat{\psi}_1(P)(a', y)}{dP f_{a', y}} &= y \frac{1}{P f_{a'}}, \\
    \frac{d\hat{\psi}_1(P)(a')}{dP f_{a'}} &= -\sum_{y \in \mathcal{Y}} y \frac{P f_{a', y}}{(Pf_{a'})^2}, \\
\end{align}
Therefore, by the functional delta method,
\begin{align}
    \sum_{f \in \mathcal{F}} \frac{d\hat{\psi}_1(P)(a')}{dPf} f(o) =& \frac{\mathbb{I}[A = a']}{P(A = a')}(y - \E_P[Y \mid A = a']). 
\end{align}
which is the EIF stated in the theorem. The proof of Theorem~\ref{theorem:von-mises} establishes that, for this EIF, the remainder term of the von-Mises expansion of $\psi_1(P)(a')$ is second-order, which allows us to conclude that this is the general EIF for $\psi_1(P)(a')$. 

A similar procedure can be used to find the EIF of $\psi_2(P)(a')$. Write the nonparametric MLE of $\psi_2(P)(a')$ as
\begin{align}
    \hat{\psi}_2(P_n)(a') = \sum_{w \in \mathcal{W}} \frac{P_n f_{w,a'}}{P_n f_{a'}} \left[ \sum_{z \in \mathcal{Z}} P_{Z\mid W}(z, w) \left[ \sum_{y \in \mathcal{Y}} \frac{y P_n f_{w,z,y}}{P_n f_{w,z}} \right] \right]
\end{align}
where
\begin{align}
    f_a(O) &= \mathbb{I}[A = a], \\
    f_{w,a}(O) &= \mathbb{I}[W = w, A = a], \\
    f_{w,z}(O) &= \mathbb{I}[W = w, A = z], \\
    f_{w,z,y}(O) &= \mathbb{I}[Y = y, A = z, W = w]. 
\end{align}
The derivatives of $\hat{\psi}_2(P)(a')$ with respect to each $f \in \mathcal{F}$ are given by
\begin{align}
    \frac{d\hat{\psi}_2(P)(a')}{dP f_{a'}} &= -\sum_{w \in \mathcal{W}} \frac{P f_{w,a'}}{(P f_{a'})^2} \left[ \sum_{z \in \mathcal{Z}} P_{Z\mid W}(z, w) \left[ \sum_{y \in \mathcal{Y}} \frac{y P f_{w,z,y}}{P f_{w,z}} \right] \right], \\
    \frac{d\hat{\psi}_2(P)(a')}{dPf_{w,a'}} &= \frac{1}{P f_{a'}} \sum_{z \in \mathcal{Z}} P_{Z\mid W}(z, w) \left[ \sum_{y \in \mathcal{Y}} \frac{y P f_{y,z,w}}{P f_{w, z}} \right], \\
    \frac{d\hat{\psi}_2(P)(a')}{dP f_{w, z}} &= -\frac{P f_{w,a'}}{P f_{a'}} \left[ P_{Z\mid W}(z, w) \left[ \sum_{y \in \mathcal{Y}} \frac{y P f_{w,z,y}}{(Pf_{w,z})^2} \right] \right], \\
    \frac{d\hat{\psi}_2(P)(a')}{dPf_{w,z,y}} &= \frac{P f_{w,a'}}{P f_{a'}} \left[ P_{Z\mid W}(z, w) \frac{y}{P f_{w, z}} \right]
\end{align}
Note that
\begin{align}
    \sum_{f \in \mathcal{F} } \frac{d\hat{\psi}_2(P)}{dPf} Pf = 0.
\end{align}
Then 
\begin{align}
    \sum_{f \in \mathcal{F}} \frac{d\hat{\psi}_2(P)(a')}{dPf} f(o) =& \frac{1}{P(A = a')} \Bigg\{
    - \mathbb{I}[a = a'] \psi_2(P)(a') \\
    &+ \mathbb{I}[a = a'] \sum_{z \in \mathcal{Z}} P_{Z \mid W}(z, w) \bar{Q}_P(z, w) \\
    &- \frac{P(W = w, A = a')}{P(W = w, A = a)} P_{Z \mid W}(z, w) \bar{Q}_P(z, w) \\
    &+ \frac{P(W = w, A = a')}{P(W = w, A = a)} P_{Z \mid W}(z, w) y \\
    =& \frac{1}{P(A = a)} \Bigg\{\frac{\g_P(a', w)}{\g_P(a, w)} P_{Z \mid W}(z, w) \left( y - \bar{Q}_P(a, w) \right) \\ 
    &+ \mathbb{I}[a = a'] \left( \sum_{z \in \mathcal{Z}} P_{Z \mid W}(z, w) \bar{Q}_P(z, w) - \psi_2(P)(a') \right) \Bigg \}
\end{align}
which is the EIF stated in the theorem, and yields a remainder term that is second-order (Theorem~\ref{theorem:von-mises}). 

The parameters $\psi_{\excessrisk}(P)(a')$ and $\psi_{\smr}(P)(a')$ are differentiable functions of $\psi_1(P)(a')$ and $\psi_2(P)(a')$ (assuming that $\psi_2(P)(a') \neq 0$, for the case of $\psi_{\smr}(P)(a')$). Therefore the EIFs for $\psi_{\excessrisk}(P)(a')$ and $\psi_{\smr}(P)(a')$ follow by simple Delta-method arguments.

\subsection{Proof of Theorem~\ref{theorem:von-mises}}
\begin{proof}
For $\psi_{1}(a')(P)$:
\begin{align}
    R_{1,a'}(P, P_0) &= \psi_1(P)(a') - \psi_1(P_0)(a') + \E_{P_0}[D_{1, a'}(P)(O)] \\
    &= \E_P[Y \mid A = a'] - \E_{P_0}[Y \mid A = a'] + \E_{P_0}\left[ \frac{\mathbb{I}[a = a']}{P(A = a')} \left\{ y - \E_P[Y \mid A = a'] \right\} \right] \\
    &= \E_{P_0}\left[ \left(\frac{P_0(A = a')}{P(A = a')} - 1\right)(\E_{P_0}[Y \mid A = a'] - \E_P[Y \mid A = a'] ) \right] \\
    &= \E_{P_0}\left[ \left(\frac{P_0(A = a')}{P(A = a')} - 1\right)(\Qa_{P_0}(a') - \Qa_P(a') ) \right].
\end{align}
For $\psi_{2}(a')(P)$: 
\begin{align}
    R_{2,a'}(P, P_0) &= \psi_2(P)(a') - \psi_2(P_0)(a') + \E_{P_0}\left[ D_{2,a'}(P)(O) \right] \\
    &= \psi_2(P)(a') - \psi_2(P_0)(a') + \E_{P_0}\left[ \frac{1}{P(A = a')} \left\{ \g_P(a', W) \left( y - \Qw_P(W) \right) + \mathbb{I}[A = a'] \Qw_P(W)\right\}  - \psi_2(P)(a') \right] \\
    &= \E_{P_0}\left[ \frac{1}{P(A = a')} \left\{ \g_P(a', W) \left( \Qw_{P_0}(W) - \Qw_P(W) \right) + \mathbb{I}[A = a']\Qw_P(W) \right\} - \frac{\mathbb{I}[A = a']}{P_0(A = a')} \Qw_{P_0}(W) \right] \\
    &= \E_{P_0}\left[ \frac{1}{P(A = a')} \left(\g_P(a', W) - g_{P_0}(a', W) \right) \left( \Qw_{P_0}(W) - \Qw_P(W) \right) \right]. 
\end{align}

\end{proof}

\section{Proof of Theorem~\ref{theorem:tmle-normality}}
\begin{proof}
For $\phi^{\tmle}(a')$, write the decomposition
\begin{align}
    \phi^{\tmle}(a') - \phi(P_0)(a') =& -P_n\left[ D_{a'}(P_n)(a') \right] \\
    &+ (P_n - P_0)\left[ D_{a'}(P_0) \right] \\
    &+ (P_n - P_0)\left[ D_{a'}(P_n) - D_{a'}(P_0) \right] \\
    &+ R_{a'}(P, P_0).
\end{align}
The first line is $o_P(n^{-1/2})$ by construction of the TMLE fluctuation. The third line is $o_P(n^{-1/2})$ due to cross-fitting. The fourth line is $o_P(n^{-1/2})$ by assumption. By the Central Limit Theorem,
\begin{align}
    \sqrt{n}(P_n - P_0)[D_{a'}(P_0)] \leadsto N(0, D_{a'}(P_0)^2).
\end{align}
The result for $\psi_1^\tmle(a')$ and $\psi_2^{\tmle}(a')$ follows identically. The result for $\psi^{\tmle}_{\excessrisk}$ and $\psi^{\tmle}_{\smr}(a')$ follows by the Delta method. 
\end{proof}
\end{appendices}

\end{document}